\documentclass[aps,prl,twocolumn,superscriptaddress,reprint,showkeys]{revtex4-2}
\usepackage{graphicx}
\usepackage{amssymb, amsmath}
\usepackage{booktabs}
\usepackage[usenames]{color}
\usepackage{bm}
\usepackage{multirow}
\usepackage{dcolumn}
\usepackage{hyperref}
\usepackage{enumerate}
\usepackage[mathlines]{lineno}
\usepackage{textcomp}
\hypersetup{
    colorlinks=true,
    linkcolor=blue,
    filecolor=gray,      
    urlcolor=blue,
    citecolor=blue,
}

\renewcommand{\color}[1]{} 

\begin{document}

\title{Real Space Visualization of Order-Disorder Transition in BaTiO$_{3}$}

\newcommand{\harvard}{ The Rowland Institute at Harvard, Harvard University, Cambridge, MA, 02138, USA}
\newcommand{\bst}{ Department of Physics, University of Science and Technology Beijing, Beijing, China}
\newcommand{\tsinghua}{State Key Laboratory of Low Dimensional Quantum Physics and Department of Physics, Tsinghua University, Beijing, China}
\newcommand{\bit}{School of Materials Science and Engineering, Beijing Institute of Technology, Beijing, China}

\affiliation{\harvard}
\affiliation{\bst}
\affiliation{\tsinghua}
\affiliation{\bit}

\author{Yang Zhang}
\email{yzhang6@fas.harvard.edu}
\affiliation{\harvard}

\author{Xiaoming Shi}
\affiliation{\bst}

\author{Suk Hyun Sung}
\affiliation{\harvard}

\author{Cong Li}
\affiliation{\tsinghua}

\author{Houbing Huang}
\affiliation{\bit}

\author{Pu Yu}
\affiliation{\tsinghua}

\author{Ismail El Baggari}
\email{ielbaggari@fas.harvard.edu}
\affiliation{\harvard}

\begin{abstract}
Ferroelectricity in BaTiO$_{3}$ was observed nearly eighty years ago, but the mechanism underlying its ferroelectric-paraelectric phase transition remains elusive.
The order-disorder transition has been recognized as playing a critical role, however, the precise nature of the order parameter still remains under scrutiny, including the local dipole direction and the correlations above and below the Curie temperature.
Using \textit{in situ} scanning transmission electron microscopy, we directly map polar displacements in BaTiO$_{3}$ across the ferroelectric-paraelectric phase transition, providing atomistic insights into a order-disorder mechanism.
Atomic tracking reveals finite polar Ti displacements in the paraelectric phase where they manifest as random polar nanoregions.
The displacements align along $<$111$>$ direction in both the ferroelectric and paraelectric phases.
The paraelectric-ferroelectric transition emerges from real-space correlations of the $<$111$>$ polar Ti displacements.
Our direct visualizations provides atomic insights into the order-disorder mechanism in the ferroelectric-paraelectric transition of BaTiO$_{3}$.
\end{abstract}


\maketitle

\section*{I. Introduction}

Structural phase transitions in materials involve changes in symmetry that are tied to functional properties.
The simplest structural phase transition involves a displacive mechanism, whereby a softening displacement mode (phonon) freezes into a static pattern.
An alternative mechanism is order-disorder, where the transition involves the evolution of correlations of preformed clusters of order. 
Ferroelectric phase transitions often follow one of these mechanisms \cite{lines2001principles}.
For instance, the phase transition of PbTiO$_{3}$ is associated with the coherent displacement of Ti and Pb atoms along $<$100$>$ relative to their positions in the high-symmetry phase \cite{dove1997theory}.
Conversely, NaNO$_{2}$ undergoes a phase transition involving an emergent ordering of N$^{3+}$ and O$^{2-}$ ions, generating long-range dipoles in the ferroelectric phase \cite{horiuchi2008organic}. 

BaTiO$_{3}$ is the first pervoskite transition-metal oxide identified to exhibit ferroelectricity \cite{von1946high}, which originates from the off-center Ti displacements relative to the center of TiO$_{6}$ octahedron ($\mathbf{\Delta}_{\text{Ti}}$).
Owing to its low switching barrier and polymorphic phases, BaTiO$_{3}$ has been extensively studied across a wide range of fields, including logic-in-memory \cite{garcia2009giant, khan2020future}, nanogenerators \cite{park2010piezoelectric, park2012flexible}, energy storage \cite{pan2021ultrahigh, zhang2024ultrahigh} and photonic devices \cite{wessels2007ferroelectric, cao2021barium}.
Despite its robust ferroelectricity and widespread applications, the mechanism underlying the \textit{FE} - \textit{PE} phase transition remains elusive.
In its bulk crystal form, BaTiO$_{3}$ has a sequence of structural transitions, spanning rhombohedral ($<$180 K), orthorhombic (180 K$-$280 K)), tetragonal (280 K$-$390 K), and cubic ($>$390 K) phases \cite{jona1962ferroelectric}.
The ferroelectric (\textit{FE}) - paraelectric (\textit{PE}) phase transition coincides with the tetragonal-cubic structural transition (the \textit{FE} and \textit{PE} correspond to tetragonal and cubic phase respectively), occurring at the Curie temperature, \textit{T$_{c}$}, of around 393 K \cite{megaw1947temperature}. 
The displacive model attributes this transition to the absence of $\mathbf{\Delta}_{\text{Ti}}$ in the \textit{PE} phase, where both the global and local dipole moment vanish \cite{cochran1960crystal,shirane1967soft,vogt1982soft} (left panel of Fig. 1A).
Conversely, the order-disorder (O-D) model posits that $\mathbf{\Delta}_{\text{Ti}}$ persist locally in the \textit{PE} phase but with no long-range correlations between them (right panel of Fig. 1A)
\cite{comes1968chain,blinc1972dynamics,
chaves1976thermodynamics,zhong1994phase,pirc2004off}. 
\textcolor{red}{Although the O-D model has been recognized as playing a critical role in BaTiO$_{3}$ through various studies spanning Raman spectroscopy, X-ray scattering, neutron scattering, infrared reflectivity, second harmonic generation and convergent-beam electron diffraction \cite{quittet1973temperature,ravel1998local,zalar2003nmr,hlinka2008coexistence,pugachev2012broken, tsuda2012nanoscale, tsuda2015two, tsuda2016direct, shao2017nanoscale}, the precise structure in the PE phase, as well as the evolution of local correlations of polar displacements in real space remains elusive, persisting as a subject of debate for over 50 years \cite{comes1968chain,comes1970desordre,itoh1985crystal,ravel1998local,zalar2003nmr,stern2004character,zalar2005nmr,nakatani2016variable}. }

Here we present real-space visualizations of ferroelectric order parameters across the \textit{FE}-\textit{PE} phase transition in BaTiO$_{3}$, uncovering the atomistic mechanism underlying the O-D model.
Using \textit{in situ} scanning transmission electron microscopy (\textit{in situ} STEM), we measure finite $\mathbf{\Delta}_{\text{Ti}}$ that persists in the \textit{PE} phase, determine the $<$111$>$ direction of $\mathbf{\Delta}_{\text{Ti}}$, and elucidate an evolution of the real-space correlations of $\mathbf{\Delta}_{\text{Ti}}$ across the \textit{PE}-\textit{FE} transition.
Our atomic tracking provides critical insights into the phase transition within the framework of the order-disorder model in BaTiO$_{3}$.
Furthermore, it establishes a direct link between real-space correlations and macroscopic signatures such as diffuse lines observed in reciprocal space studies \cite{ravy2007high,page2010probing,senn2016emergence,pasciak2018dynamic}.

\begin{figure}
    \centering
    \includegraphics[width=\linewidth]{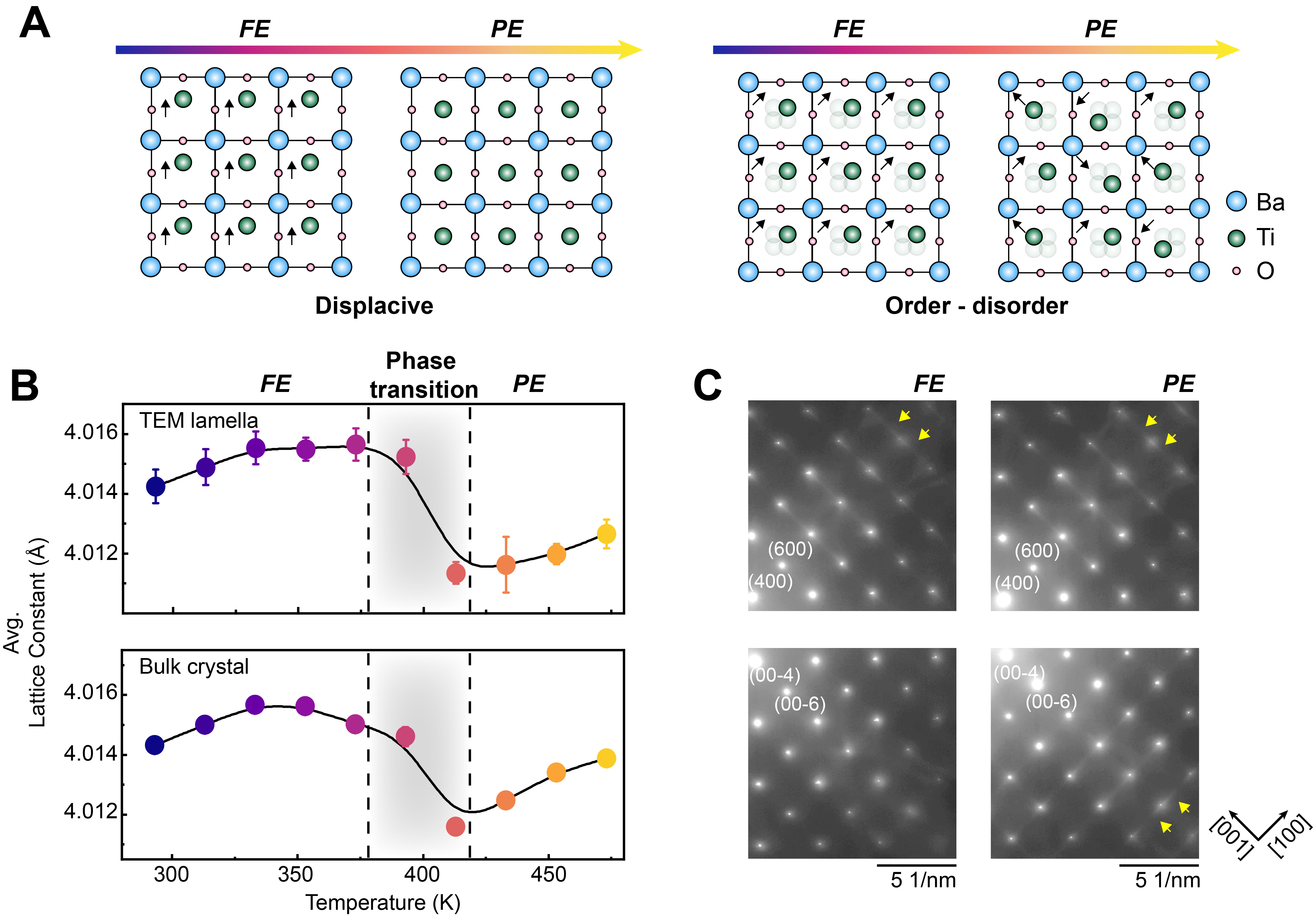}
    \caption{Ferroelectric (\textit{FE}) - paraelectric (\textit{PE}) phase transition of BaTiO$_{3}$ reproduced by \textit{in situ} STEM.
    (A) Schematic graphs showing the displacive model (left panel) and order-disorder model (right panel) of the \textit{PE}-\textit{PE} phase transition.
    The arrow represents the Ti displacement.
    The transparent atom highlights the occupational distribution in order-disorder case.
    \textcolor{red}{(B) The averaged lattice constant measured from single domain in ADF-STEM image of TEM lamella (upper panel) and X-ray diffraction (lower panel) of bulk crystal at different temperatures. }
    The error bar comes from the mean absolute error of the normal distribution fitting.
    The dashed line and gray shadow determine the phase transition region.
    (C) Zoom-in diffraction patterns collected from \textit{FE} (left panel) and \textit{PE} (right panel) phases.
    The projection is [010] zone axis.
    The yellow arrows highlight the diffuse intensity.
    }
    \label{fig1}
\end{figure} 

\section*{II. Results}

\begin{center}
\textbf{A. Finite $\mathbf{\Delta}_{\text{Ti}}$ in paraelectric phase}
\end{center}

We first perform measurements of the lattice constant across the \textit{FE}-\textit{PE} transition in an thin, electron transparent BaTiO$_{3}$ sample for transmission electron microscopy.
As shown in Fig. 1B, the lattice constant obtained from annular dark-field (ADF-STEM) images exhibits a drop near 390 K, consistent with X-ray diffraction (XRD) measurements of the bulk crystal (lower panel of Fig. 1B and raw data is shown in Fig. S1). 
The negative thermal expansion coincides with the \textit{FE}-\textit{PE} phase transition temperature of BaTiO$_{3}$ \cite{shirane1952transition, chen2015negative}.
The agreement between transmission electron microscopy and XRD demonstrates that the phase transition behavior in the electron transparent sample matches bulk BaTiO$_{3}$.

Across the the Curie temperature, the selected area electron diffraction (SAED) patterns reveal some differences between the \textit{FE} and \textit{PE} phases.  
Figure 1C shows SAED patterns collected along the [010] zone axis of BaTiO$_{3}$ at 293 K (\textit{FE}) and 473 K (\textit{PE}), respectively (raw data is shown in Fig. S2).
In addition to sharp Bragg spots, we observe diffuse intensity in both phases (highlighted by yellow arrows), indicating the presence of correlated disorder \cite{keen2015crystallography}.
Diffuse intensity lines are prominent at high index spots, suggesting the contribution from positional disorder instead of chemical disorder (see SI notes and Fig. S3 for more details).
In the \textit{FE} phase, diffuse intensity lines are present only along the [001] direction, whereas in the \textit{PE} phase, they appear along both the [001] and [100] directions, indicating a transition from anisotropic to isotropic local correlations \cite{ravy2007high,page2010probing,senn2016emergence,pasciak2018dynamic}.

To understand the microscopic structure of this disorder, we quantify $\mathbf{\Delta}_{\text{Ti}}$ in real space, which can be mapped quantitatively for each unit cell through fitting atomic positions in ADF-STEM images (see SI notes).
The upper panel of Fig. 2A shows an ADF-STEM image of BaTiO$_3$ at 473 K, where the Ba atomic columns appear brighter than Ti columns due to their larger atomic number.
Figure 2B presents a large field-of-viewed ADF-STEM image overlaid with $\mathbf{\Delta}_{\text{Ti}}$ at 473 K (\textit{PE}) and 293 K (\textit{FE}). 
The color represents the direction of $\mathbf{\Delta}_{\text{Ti}}$ and transparency reflects the amplitude of the displacement.
Even above T$_{c}$, the \textit{PE} phase exhibits short-range polar clusters with finite $\mathbf{\Delta}_{\text{Ti}}$, close to 9 pm.
Displacement magnitudes in the PE phase are larger than the 4 pm precision of the measurement (Fig. S4).
The ADF-STEM image reveals no apparent crystalline defects or boundaries that would pin polar displacements in the \textit{PE} phase.
\textcolor{red}{The possible influence of oxygen vacancies generated at high temperatures is also excluded (Fig. S5).}
Thus, these visualizations provide direct, real-space evidence for finite $\mathbf{\Delta}_{\text{Ti}}$ well into the \textit{PE} phase.

\begin{center}
\textbf{B. $<$111$>$ displacement direction}
\end{center}

We next address the direction of $\mathbf{\Delta}_{\text{Ti}}$; 
whether it inherits $<$001$>$ tetragonal displacements or $<$111$>$ rhombohedral displacements.
To do so, we directly quantify the projected shift direction (\textit{$\phi$}), as defined in the right panel of Fig. 2A.
\textcolor{red}{In the \textit{FE} phase, long-range order starts to emerge when compared with PE phase (lower panel of Fig. 2B),} but the \textit{$\phi$} does not align well with either the $<$001$>$ (\textit{$\phi$}=$0^\circ$) or $<$111$>$ (\textit{$\phi$}=$45^\circ$) projected directions (left panel of Fig. 2C).
Instead, \textit{$\phi$} in the \textit{FE} phase aligns close to $<$001$>$ projected direction (right panel of Fig. 2C).
Additionally, when we tracked the evolution of \textit{$\phi$} across the \textit{FE}-\textit{PE} phase transition, we find large changes within the phase transition temperature region (Fig. 2D and raw data is shown in Fig. S6).
This evolution is also reproducible across regions with different domain configurations (Figs. S7-S8), excluding the influence of domain boundaries.

\begin{figure}
    \centering
    \includegraphics[width=\linewidth]{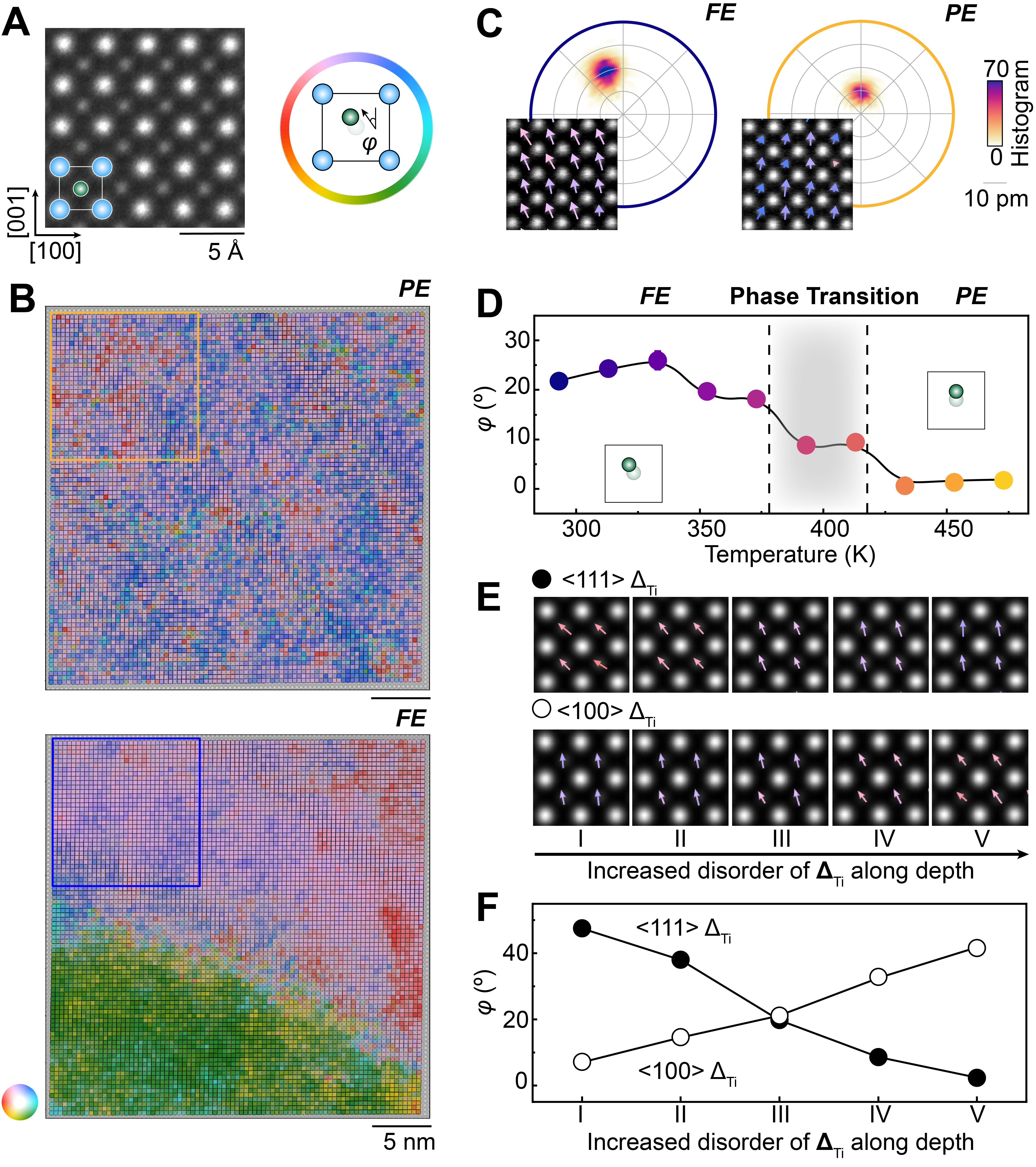}
    \caption{ Finite $\mathbf{\Delta}_{\text{Ti}}$ in \textit{PE} phase and evidence of $<$111$>$ displacements.
    (A) Left panel: Zoomed-in ADF-STEM image of BaTiO$_{3}$ overlapped with an atomic model showing Ba (blue) and Ti (green).
    The projection is along the [010] direction.
    Right panel: definition of the direction of $\mathbf{\Delta}_{\text{Ti}}$ (\textit{$\phi$}).
    (B) Real-space map of $\mathbf{\Delta}_{\text{Ti}}$ in \textit{FE} and \textit{PE} phase.
    The color and transparent represents the direction and amplitude, respectively.
    (C) The polar histogram of $\mathbf{\Delta}_{\text{Ti}}$ in \textit{FE} and \textit{PE} phase.
    The radius of the polar histogram plot 40 pm.
    The inset shows zoomed-in ADF-STEM image overlapped with measured $\mathbf{\Delta}_{\text{Ti}}$ in the \textit{FE} and \textit{PE} phase.
    (D) Evolution of \textit{$\phi$} with temperature.
    The inset shows the schematic graph of the changes between \textit{FE} and \textit{PE} phase.
    The black dashed line and gray shadow highlight the phase transition region.
    The plot summarizes data from the region marked by a rectangle in (B) to exclude the effect of \textit{FE} domain boundary.
    (E) Multislice image simulations and mapping of projected $\mathbf{\Delta}_{\text{Ti}}$ with increasing disorder (I-V) along the depth (electron beam imaging direction).
    The upper and lower panel represent \textbf{$\Delta_{Ti}$} along the $<$111$>$ and $<$100$>$ direction, respectively.
    (F) \textit{$\phi$} of the projected displacements from multislice simulations.
    }
    \label{fig2}
\end{figure}

We show that this abnormal alignment and change of \textit{$\phi$} represents the distinct disorder configurations $<$111$>$ rhombohedral displacements in both the \textit{FE} and \textit{PE} phases. 
ADF-STEM is a projection imaging method along a column of atoms.
\textcolor{red}{To simplify, we only consider the disorder in the displacements along the depth direction instead of 3-dimensional disorder to interpret changes in the projected displacements (see SI notes).}
Multislice image simulations can be used to study the effect of disorder in projection.
We construct supercells with random stackings of $<$111$>$ and $<$001$>$ $\mathbf{\Delta}_{\text{Ti}}$ variants along the depth direction (Figs. S8-S9).
Figures 2E-F summarize the dependence of \textit{$\phi$} on the disorder of $\mathbf{\Delta}_{\text{Ti}}$ along the depth direction. 
The \textit{$\phi$} determined in the experiment is reproducible in the case of rhombohedral-like $<$111$>$ displacements, which shows a monotonic decrease from $45^\circ$ to $0^\circ$ when we increase disorder.
In contrast, the tetragonal-like $<$100$>$ displacement direction shows the opposite trend and inconsistent with the experimental data in Fig. 2D.
Therefore, the FE phase, nominally in the "tetragonal" phase below T$_{c}$, comprises rhombohedral displacements which, in projection STEM, \textcolor{red}{manifest as \textit{$\phi$} that increases as temperature/disorder decreases.}
The temperature trend (Fig. 2D) and simulations of disorder(Fig. 2F) further indicate that the PE phase inherits $<$111$>$ displacements but with increased disorder.

\begin{figure}
    \centering
    \includegraphics[width=\linewidth]{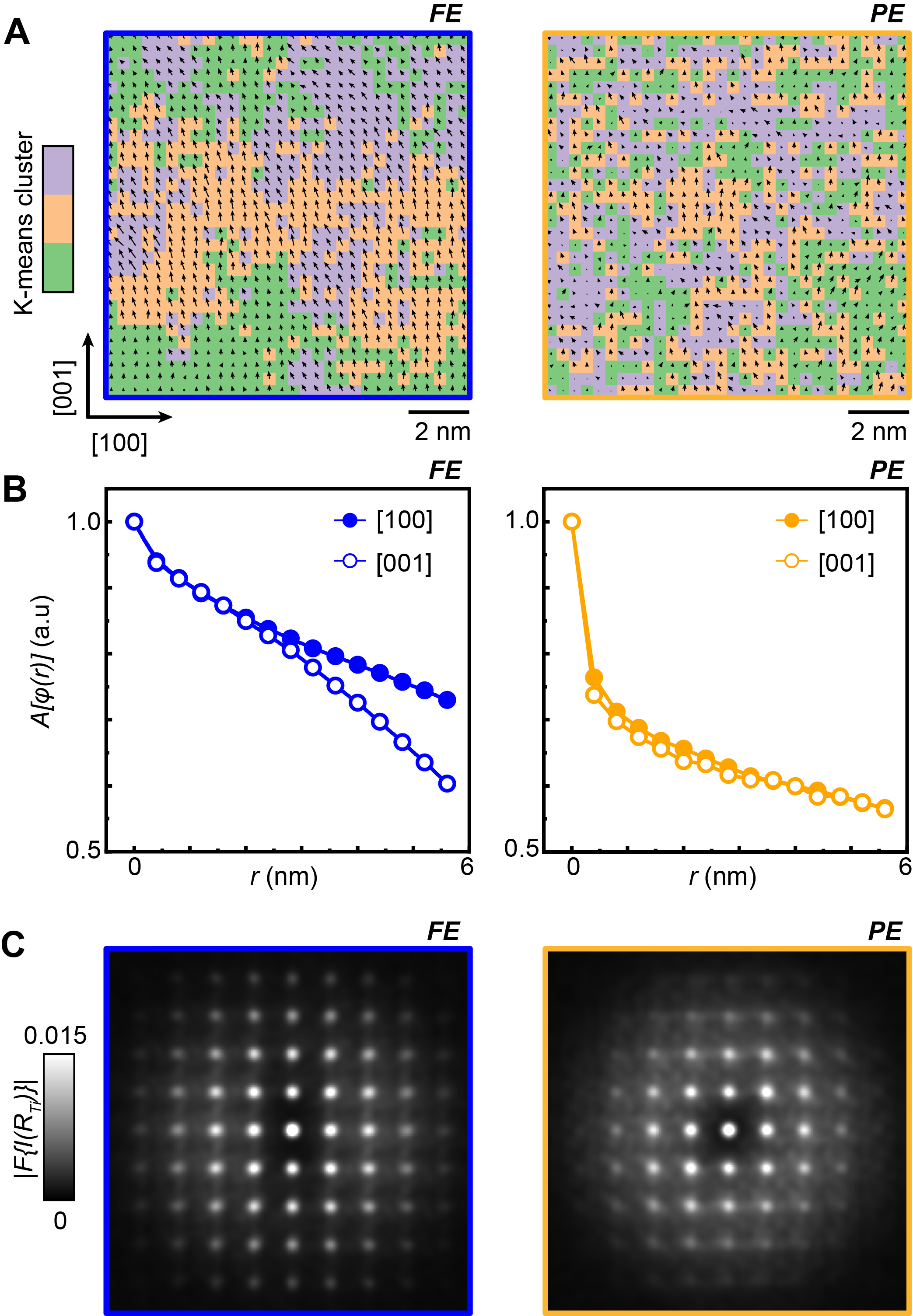}
    \caption{ Real space correlations of polar displacements $\mathbf{\Delta}_{Ti}$.
    (A) K-means clustering of the region marked by rectangle in Fig. 2B, see details in SI notes.
    The color represents individual clusters of correlated displacements.
    The arrow represent $\mathbf{\Delta}_{Ti}$, with the size and length encoding the amplitude and direction, respectively
    (B) Autocorrelation function of $\mathbf{\Delta}_{Ti}$ (\textit{$A[\phi(\mathbf{r})]$}) extracted along [100] (solid circle) and [001] (hollow circle) directions.
    (C) Fourier transform of 2D Ti positions in ferroelectric (left) and paraelectric (right). Anisotropy in the diffuse intensity is evident.
    }
    \label{fig3}
\end{figure}

\begin{figure*}
    \centering
    \includegraphics[width=\linewidth]{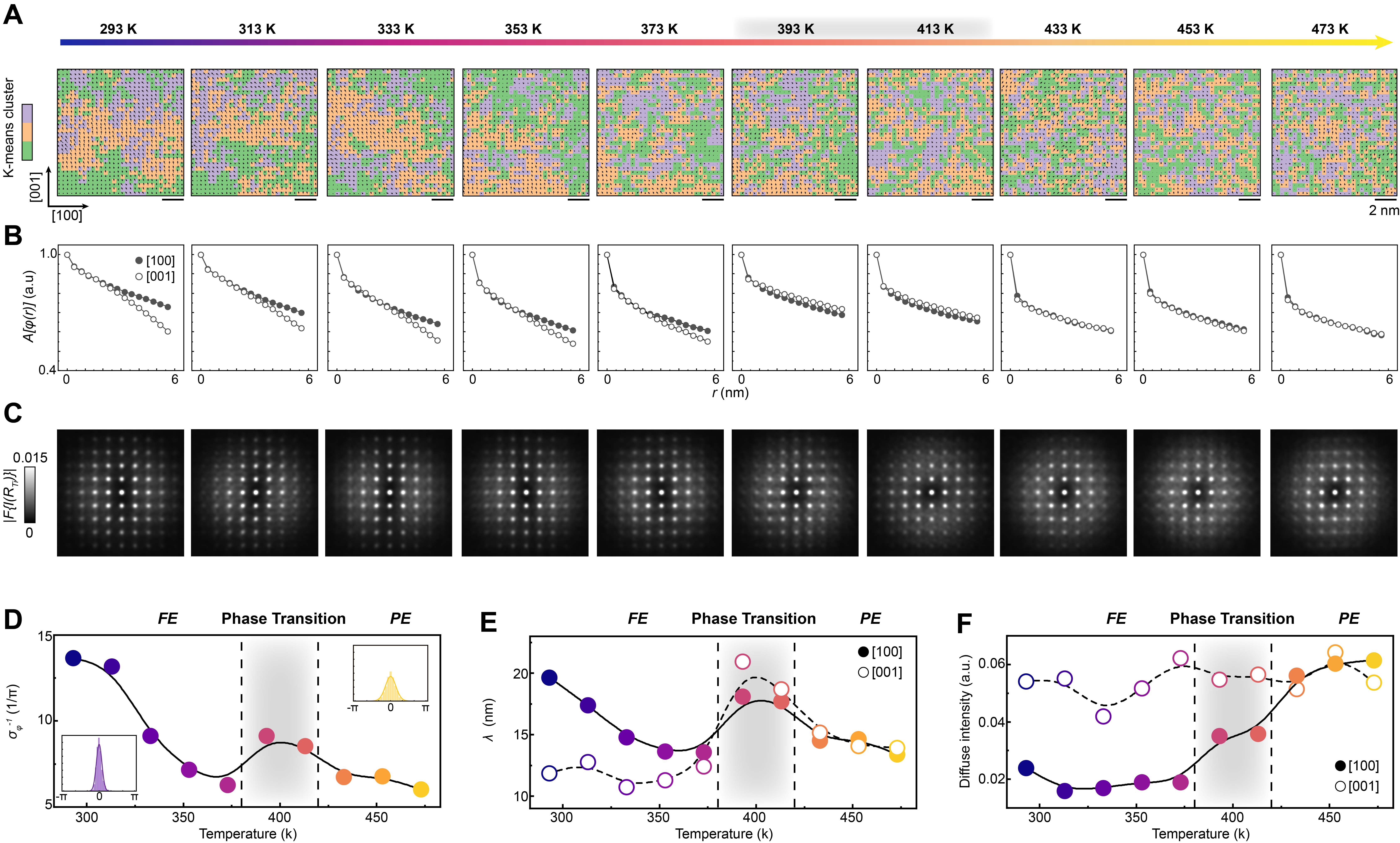}
    \caption{ Evolution of real space correlations of $\mathbf{\Delta}_{\text{Ti}}$ across temperatures.
    (A) K-means clusters, (B) autocorrelation function of $\mathbf{\Delta}_{\text{Ti}}$ and (C) Fourier transform result of 2D Ti positions measured across temperatures.
    (D) Standard deviation of displacement direction, \textit{$\sigma_{\phi}^{-1}$}, at different temperatures. 
    The inset shows the histogram at 293 K and 473 K, respectively.
    The mean value was set to 0.
    (E) Correlation length, (\textit{$\lambda$}), extracted from  $A[\phi(\mathbf{r})]$ at different temperatures.
    The hollow and solid circle represents the (\textit{$\lambda$}) along [001] and [100] direction, respectively.
    (F) Temperature-dependent Diffuse intensity in the Fourier transform.
    The diffuse intensity are normalized to the the (200) and (002) Bragg peaks.
    The hollow and solid circle represents the [001] and [100] direction, respectively.
    The black dashed line and gray shadow highlights the phase transition region.
    }
    \label{fig4}
\end{figure*}

\begin{center}
\textbf{C. Emergence of $\mathbf{\Delta}_{\text{Ti}}$ correlations in real space}
\end{center}

To elucidate local correlations between nearby Ti sites and their relationship to diffuse intensity in diffraction, we visualize fluctuations in $\mathbf{\Delta}_{\text{Ti}}$.
Clustering analysis was performed to enhance the visibility of these fluctuations (see SI notes for details).
Clustering identifies nanoscale regions (colors) with different displacement directions/amplitudes (arrows). 
Despite having large ferroelectric domains, the FE phase still shows local fluctuations within a single domain, as captured by the clusters (Fig. 3A).
In the \textit{PE} phase, the clusters are smaller and more randomly distributed.

To quantify the real-space correlations, we calculated the 2D autocorrelation function of $\mathbf{\Delta}_{\text{Ti}}$ (\textit{$A[\phi(\mathbf{r})]$}), and extracted its decay behavior along two orthogonal directions, [100] and [001] (see SI notes for more details).
As shown in Fig. 3B, the decay of \textit{$A[\phi(\mathbf{r})]$} encodes the degree of correlations between $\mathbf{\Delta}_{\text{Ti}}$, with a faster decay indicating weaker correlations.
In the \textit{FE} phase, an anisotropic correlation is evident, with slower decay along [100] indicating stronger correlation compared to [001].
In the \textit{PE} phase, \textit{$A[\phi(\mathbf{r})]$} decays more rapidly and appears more isotropic.

The effect of local $\mathbf{\Delta}_{\text{Ti}}$ correlations can further be linked to diffuse intensity in reciprocal space, through Fourier transformation of the 2D Ti sublattice (see SI notes and Figs. S10-S11 for details).
As summarized in Fig. 3C, the Fourier transform of Ti positions in the \textit{FE} phase exhibits diffuse intensity along the [001] direction, consistent with anisotropic correlations revealed by \textit{$A[\phi(\mathbf{r})]$}.
In the \textit{PE} phase, however, diffuse intensity is observed along both orthogonal directions, which indicates a transition to isotropic behavior, also in agreement with \textit{$A[\phi(\mathbf{r})]$}.

Detailed in situ heating in the same field of view helps track the evolution of these parameters.
The temperature-dependent behaviors of the fluctuations, correlation function and diffuse intensity anisotropy are summarized in Fig. 4.
Fluctuations become more significant across the phase transition (Fig. 4A), further quantified by the standard deviation of \textit{$\phi$}.
As exhibited in Fig. 4D, \textit{$\sigma_{\phi}$} grows with increasing temperature in both \textit{FE} and \textit{PE} phases (raw data is shown in Fig. S14).
Figure 4B shows the \textit{$A[\phi(\mathbf{r})]$} along [100] and [001] direction at different temperatures, confirming that anisotropy vanishes as the system transits through the phase transition region.
\textcolor{red}{Quantification of correlation lengths (\textit{$\lambda$}, see SI notes for more details) along two directions (Fig. 4E) shows the same trend.
The difference of \textit{$\lambda$} between two orthogonal directions begins to disappear when BaTiO$_{3}$ crosses the FE-PE transition.}
Likewise, diffuse intensity in Fourier transform follows the same trend.
As displayed in Figs. 4C and F, diffuse intensity becomes isotropic across the phase transition region, as revealed through linecuts along the [001] and [100] directions (Fig. S15).
The evolution of diffuse intensity in Fourier transform is also consistent with that observed in diffraction pattern (Fig. S16).
These three parameters show a high degree of consistency, demonstrating that local correlations of $\mathbf{\Delta}_{\text{Ti}}$ dictates the \textit{FE}-\textit{PE} phase transition.

\begin{figure}
 \centering
    \includegraphics[width=\linewidth]{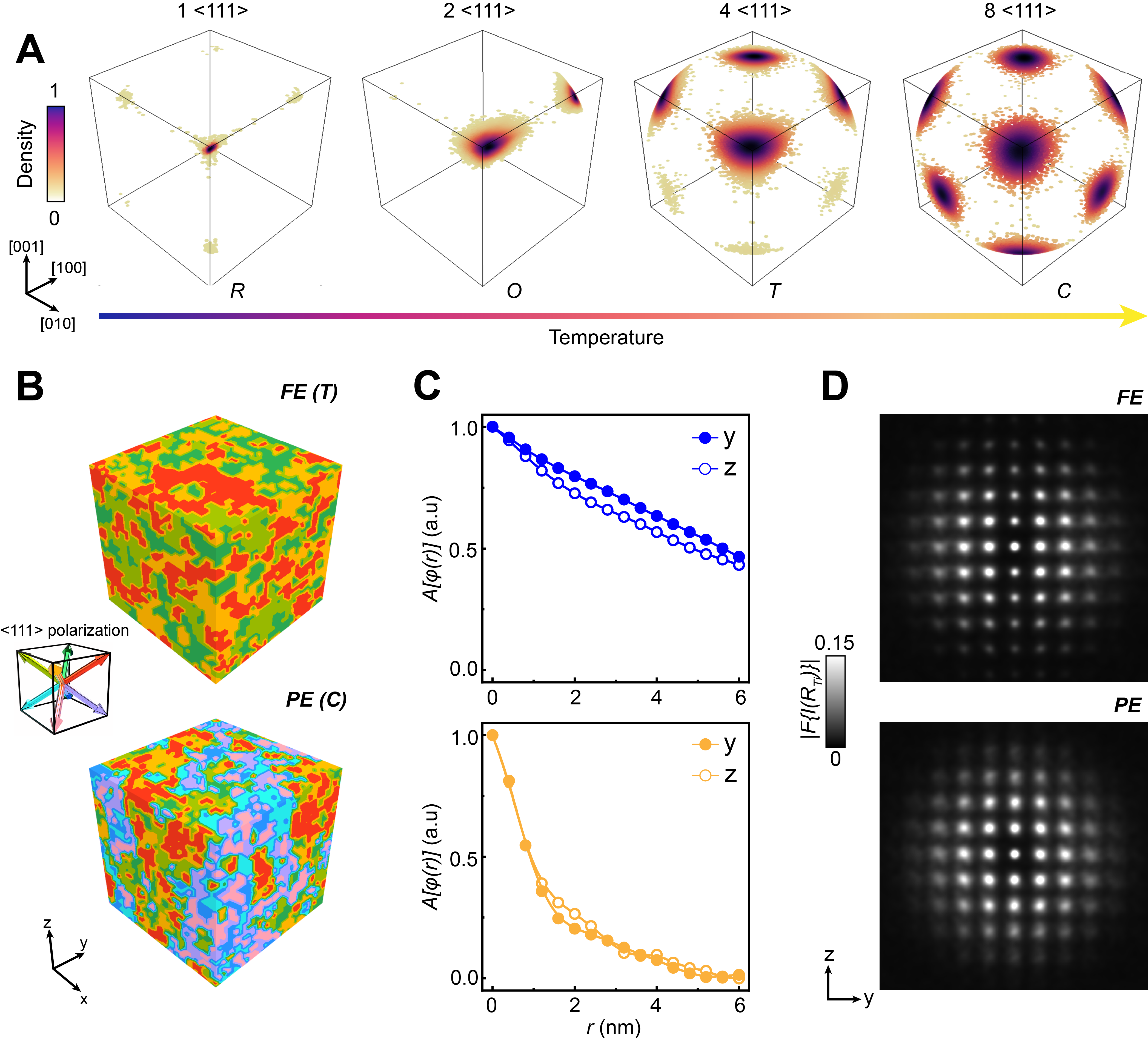}
    \caption{Phase field simulations of a order-disorder mechanism in BaTiO$_{3}$.
    (A) Occupation of Ti displacement direction position at different temperatures collected from phase field simulations.
    The density is normalized from 0 to 1 for each temperature.
    (B) Distribution of eight $<$111$>$ $\mathbf{\Delta}_{\text{Ti}}$ within a 32$\times$32$\times$32 super cell in \textit{PE} and \textit{PE} phase.
    (C) Autocorrelation function of yz-plane projected Ti displacements (\textit{$A[\phi(\mathbf{r})]$}).
    The solid and hollow circle represents the profile extracted along the y (solid circle) and x (hollow circle) direction.
    (D) Fourier transform of 2-dimensional image reconstructed from  Ti displacements projected on y-z plane.
    The raw data of projected Ti displacements and reconstructed 2D patterns are shown in Fig. S17.
    }
    \label{fig4}
\end{figure}

\begin{center}
\textcolor{red}{\textbf{D. Capturing all phase transition in BaTiO$_{3}$ under order-disorder framework}}
\end{center}

The atomistic mechanism of O-D model we identified in BaTiO$_{3}$, involving (1) persistently finite $\mathbf{\Delta}_{\text{Ti}}$ in the \textit{PE} phase, (2) displacements along the $<$111$>$ directions, and (3) the key role of correlations of $\mathbf{\Delta}_{\text{Ti}}$, motivates a simple phase-field simulation to replicate the phase transitions in BaTiO$_{3}$.
Starting from our experimental observation, we only considered $<$111$>$-type $\mathbf{\Delta}_{Ti}$ and incorporated the interplay between correlation of $\mathbf{\Delta}_{Ti}$ and thermal fluctuations in a phase field simulation (see Materials and Methods).
\textcolor{red}{As displayed in Fig. 5A, the occupation probabilities of $<$111$>$ directions undergo three distinct changes, mirroring the typical phase transition sequence in BaTiO$_3$ (\textit{R}-\textit{O}-\textit{T}-\textit{C}), consistent with other theoretical works \cite{gigli2022thermodynamics, ovcenavsek2023dynamics, kotiuga2022microscopic}.}

We further mapped the spatial distribution of $<$111$>$ $\mathbf{\Delta}_{\text{Ti}}$ within a 32$\times$32$\times$32 supercell for both \textit{FE} and \textit{PE} phases, and projected the $\mathbf{\Delta}_{\text{Ti}}$ onto the yz-plane (Fig. S17).
As depicted in Fig. 5B, the fluctuations of $\mathbf{\Delta}_{\text{Ti}}$ are evident in both the \textit{FE} and \textit{PE} phases, with more significant fluctuations observed in the \textit{PE} phase.
To quantify the correlations, we extracted \textit{$A[\phi(\mathbf{r})]$} along y- and z-direction, and also performed Fourier transform of positional pattern, analogous to the STEM analysis.
Figures 5C shows the \textit{$A[\phi(\mathbf{r})]$} extracted along two directions in \textit{FE} and \textit{PE} phases, with decays that align with the trends found in experimental data.
There is an anisotropic decay along two directions in the \textit{FE} phase, while the \textit{PE} phase exhibits isotropic and faster decay.
Further, the Fourier transform shown in Fig. 5D reveals changes in diffuse intensity that mirror those in experiment.

\section*{III. Discussion}
In this work, we present direct real-space visualizations of ferroelectric order parameters in BaTiO$_3$ across the ferroelectric-paraelectric phase transition.
\textcolor{red}{With the help of real-space visualization, we provide new atomistic insights into the controversies of the order-disorder model in BaTiO$_{3}$.}
(i) finite $\mathbf{\Delta}_{\text{Ti}}$ are evident in the paraelectric phase.
(ii) the displacement direction of $\mathbf{\Delta}_{\text{Ti}}$ in \textit{PE} phase, inherited from the \textit{FE} phase, remains aligned along $<$111$>$ directions. 
(iii)$\mathbf{\Delta}_{\text{Ti}}$ show distinct local correlations across temperature, and their variation between anisotropy and isotropy generates diffuse intensity in reciprocal space.
All these parameters undergo significant changes across the phase transition region.
This work demonstrates the order-disorder nature of phase transition in BaTiO$_{3}$ and highlights the critical role of atomic-scale visualizations in revealing local correlation within the ordered and disordered states.
\textcolor{red}{Furthermore, our phase-field model will help determine the role of order-disorder transitions in mesoscale domains and ferroelectric-related properties of BaTiO$_{3}$ and other perovskite ferroelectrics.}

\section*{Methods}

\begin{center}
\textbf{A. Sample preparation}   
\end{center}

BaTiO$_{3}$ is a commercial single crystal (MTI Corporation).
The cross-sectional STEM specimens were prepared using standard Gallium focused ion beam (Thermo-Fisher Helios) lift-out and thinning.
The samples were thinned down using an accelerating ion voltage of 30 kV with a decreasing current from 100 pA to 40 pA, and then with a fine polishing process using an accelerating voltage of 5 kV and 2 kV, and a current of 41 pA and 23 pA.
\textcolor{red}{The thickness of our BaTiO$_{3}$ lamella was estimated to be $\sim\!30~\mathrm{nm}$ by zero-loss peak in EELS \cite{malis1988eels}.}

\begin{center}
\textbf{B. \textit{In situ} X-ray diffraction measurement} 
\end{center}

Single crystal BaTiO$_{3}$ was first fixed to the heater holder with colloidal silver to ensure good thermal contact. 
The $\theta$-$2\theta$ scanning was carried out with a high-resolution diffractometer (Smartlab, Rigaku) using monochromatic Cu $K_{\alpha\mathbf{1}}$ radiation ($\lambda$ = 1.5406 \text{\AA}) in an atmosphere of argon gas at different temperature stages.
The temperature range was from $30~^\circ\mathrm{C}$ to $250~^\circ\mathrm{C}$, while the step was $10~^\circ\mathrm{C}$. 
All temperature-dependent measurements were taken 5 minutes after the temperature had reached the set point. 
The lattice parameters of a- and c-domain were calculated from the position of the diffraction peak (002), which were determined via piecewise Gaussian fitting of both separate peaks.

\begin{center}
\textbf{C. Scanning transmission electron microscopy}
\end{center}

In situ STEM experiments were performed in an aberration-corrected microscope (Thermo-Fisher Scientific Themis Z G3) operated at 200 kV. 
The in situ heating experiment was carried out using MEMS-based heating/bias chips (DENSsolutions).
\textcolor{red}{The temperature range was from $20~^\circ\mathrm{C}$ to $200~^\circ\mathrm{C}$ with $20~^\circ\mathrm{C}$ step.}
ADF-STEM images at all temperatures were collected using a 18.9 mrad convergence angle and 30 pA probe current. 
The collection angles ranged from 68 to 200 mrad, corresponding to high-angle ADF. 
This imaging mode is mostly sensitive to atomic number and has interpretable contrast, making quantification of atomic positions reliable.  
To minimize image drift and obtain a high signal-to-noise ratio, fast-acquisition frames were collected.
Each frame was acquired with 2048$\times$2048 pixels and 100 ns dwell time.
Fifty frames in total were collected and aligned using a rigid registration method optimized for noisy image frames \cite{savitzky2018image}.
The ADF-STEM images collected at different temperatures come from the same field of view. 

\begin{center}
\textbf{D. Phase field simulation}     
\end{center}

In traditional ferroelectric phase field models, a position-dependent spontaneous polarization $\bm{P}$ and a displacement field $\bm{u}$ serve as order parameters. 
Polarization dynamics is governed by the time-dependent Ginzburg-Landau (TDGL) equation, while the corresponding equilibrium equations capture the equilibrium of stress and electric fields \cite{shi2022quantitative}:

\begin{equation}
\frac{\partial \bm{P}}{\partial t} = -L \frac{\delta F}{\delta \bm{P}}, \quad \nabla \cdot \bm{\sigma} = 0, \quad \nabla \cdot \bm{D} = \rho_f,
\end{equation}

where $L$ is a kinetic coefficient related to domain wall mobility, $F$ is the total free energy of the system, $\frac{\delta F}{\delta \bm{P}}$ is the thermodynamic driving force for polarization evolution, $\sigma_{ij}$ is the stress tensor, $\bm{D}$ is the electric displacement, $\rho_f$ is the free charge density, and $\bm{r}$ and $t$ are the spatial coordinate and time, respectively. The thermal field $\bm{E}^{\text{thermal}}$ follows \cite{yang2020domain}:

\begin{equation}
\begin{aligned}
& \, \langle E^{\text{thermal}}_i(x_1, t_1) E^{\text{thermal}}_j(x_2, t_2) \rangle = \\
& 2 k_B T L \delta(x_1 - x_2) \delta(t_1 - t_2),
\end{aligned}
\end{equation}

where $k_B$ is the Boltzmann constant and $T$ is the thermodynamic temperature. The total free energy of a bulk system can be defined as follows:

\begin{equation}
F = \int_V \left( f_{\text{Land}} + f_{\text{grad}} + f_{\text{elastic}} + f_{\text{elec}} \right) \, dV,
\end{equation}

where $F$ includes the bulk free energy $f_{\text{Land}}(\bm{P})$, domain-wall energy $f_{\text{grad}}$, elastic energy $f_{\text{elastic}}$, and electrostatic energy $f_{\text{elec}}$, with $\bm{E}$ as the applied static electric field. The corresponding energy densities are $f_{\text{Land}}, f_{\text{grad}}, f_{\text{elastic}},$ and $f_{\text{elec}}$.

The bulk free-energy density is expressed as a sixth-order polynomial expansion:

\begin{equation}
\begin{aligned}
f_{\text{Land}} = & \, a_1 (P_1^2 + P_2^2 + P_3^2) 
+ a_{11} (P_1^4 + P_2^4 + P_3^4) \\
& + a_{12} (P_1^2 P_2^2 + P_2^2 P_3^2 + P_1^2 P_3^2) \\
&+ a_{111} (P_1^6 + P_2^6 + P_3^6) \\
& + a_{112} \big[P_1^4 (P_2^2 + P_3^2) 
+ P_2^4 (P_1^2 + P_3^2) \\
& + P_3^4 (P_1^2 + P_2^2)\big]
+ a_{123} P_1^2 P_2^2 P_3^2.
\end{aligned}
\end{equation}

where a$_{1}$ to a$_{123}$ are Landau parameters.

The gradient energy density in an anisotropic system can be calculated by:

\begin{equation}
f_{\text{grad}} = \frac{1}{2} g_{ijkl} P_{i,j} P_{k,l},
\end{equation}

where $g_{ijkl}$ is the gradient energy coefficient and $P_{i,j} = \frac{\partial P_i}{\partial x_j}$. The elastic energy density is given by:

\begin{equation}
f_{\text{elastic}} = s_{ij} \sigma_{ij}^2,
\end{equation}

where $s_{ij}$ is the compliance coefficient. Here, $\sigma_{ij} = C_{ijkl} (\varepsilon_{kl} - \varepsilon_{kl}^0)$, with $C_{ijkl}$ as the stiffness tensor, $\varepsilon_{ij}$ as the strain tensor, and $\varepsilon_{ij}^0$ as the eigenstrain, defined as:

\begin{equation}
\varepsilon_{ij}^0 = Q_{ijkl} P_k P_l,
\end{equation}

where $Q_{ijkl}$ is the electrostrictive coefficient tensor. 

The electrostatic energy density $f_{\text{elec}}$ in the phase-field simulation is given by:

\begin{equation}
f_{\text{elec}} = -P_i(r) E_i(r) - \frac{1}{2} P_i(r) E_i^{\text{in}}(r),
\end{equation}

where $E_i^{\text{in}}(r)$ is the $E$-field induced by the dipole moments, and $E_i(r)$ is the applied electric field.

To reproduce the phase transition from R phase to O, T, and C phases, we use $a_1 = a_0 (T - T_c)$, with other Landau parameters fitted to represent different ferroelectric polarization phases in specific directions. 
\textcolor{red}{Our phase-field simulation is based on eight-site model of BaTiO$_{3}$.}
In our model, only the R phase potential well is introduced, with fixed Landau parameters as temperature increases, showing rhombohedral-like local displacements at all temperatures. 
The phase transition originates from competition between thermal fluctuations and nearest-neighbor (including second and third nearest neighbor) interactions.
As shown in Ref. \cite{senn2016emergence}, interactions were added to penalize dipole misalignment in first-, second-, and third-nearest neighbor sites. 
In the mesoscale phase-field model, this is simplified to anisotropic fluctuations in different directions, with the thermal field given by:

\begin{equation}
E_i^{\text{thermal}}(x, t) = \eta(x, t) \sqrt{\frac{2 k_B T \gamma}{\tau \Delta V}} - b_i,
\end{equation}

\textcolor{red}{where $\eta$ is a random vector, $\gamma$ is the damping coefficients of polarization evolution, $\Delta V$ is the volume of a grid cell, and $\tau$ is the time step.
$b_{i}$ is the reduced anisotropic temperature fluctuation field parameters. 
The term $b_{i}$ is then derived by reproducing the R-O-T-C transition within a reasonable transition temperature range.}


\textcolor{red}{The other parameters (all in SI units) are: 
\begin{center}
\begin{tabular}{l@{\hspace{2em}}r}
$a_1$     & $-3.8 \times 10^8~\mathrm{m^2 N/C^2}$ \\
$a_{11}$  & $4.8 \times 10^8~\mathrm{m^6 N/C^4}$ \\
$a_{12}$  & $1.9 \times 10^8~\mathrm{m^6 N/C^4}$ \\
$Q_{11}$  & $0.1~\mathrm{m^4/C^2}$ \\
$Q_{12}$  & $-0.034~\mathrm{m^4/C^2}$ \\
$Q_{44}$  & $0.029~\mathrm{m^4/C^2}$ \\
$s_{11}$  & $9.1 \times 10^{-12}~\mathrm{m^2/N}$ \\
$s_{12}$  & $-3.2 \times 10^{-12}~\mathrm{m^2/N}$ \\
$s_{44}$  & $8.2 \times 10^{-12}~\mathrm{m^2/N}$ \\
$\gamma$  & $0.05~\mathrm{Jm/(As^2)}$ \\
$\tau$    & $5 \times 10^{-12}~\mathrm{s}$ \\
$a_0$     & $3.4 \times 10^8~\mathrm{J/m^3}$ \\
$p_0$     & $0.5~\mathrm{C/m^2}$ \\
$b_2$     & $0.8$ \\
$b_3$     & $1.2$ \\
$g_{11}$  & $0.6$ \\
\end{tabular}
\end{center}
}

The simulation scale is $64\Delta x \times 64\Delta y \times 32\Delta z$, with grid scales $\Delta x$ and $\Delta z$ at 1 nm. 
Fourier methods were used to solve the equations. 
Open circuit electrical boundary conditions and periodic mechanical boundary conditions were adopted in the calculations.

\section{Conflict of interest}
The authors declare no conflict of interest.

\section{Acknowledgments}
Y. Z., S. H. S and I. E were supported by the Rowland Institute at Harvard.
C. L and P. Y were supported by the National Natural Science Foundation of China (Grant No. 52025024, and No. 52388201).
Focused ion beam sample preparation was performed at the Harvard University Center for Nanoscale Systems (CNS); a member of the National Nanotechnology Coordinated Infrastructure Network (NNCI), which is supported by the National Science Foundation under NSF award no. ECCS-2025158. 
Transmission electron microscopy was carried out through the use of MIT.nano's facilities.

Y. Z. initiated this work and performed sample preparation, STEM measurement and data analysis with the help of S. S. and I. E.
X. S. carried out the phase field simulation with the help of H. H.
C. L. carried out the \textit{in situ} XRD measurement under the supervision of P. Y.
Y. Z. and I. E. wrote the paper with contributions from all authors.

\bibliographystyle{ref}
\bibliography{reference}

\end{document}